\documentclass[%
 reprint,
 amsmath,amssymb,
 aps,
 prx,
]{revtex4-2}

\usepackage{graphicx}% Include figure files
\usepackage{dcolumn}% Align table columns on decimal point
\usepackage{bm}% bold math
\usepackage{multirow}%
\usepackage{amsmath,amssymb,amsfonts}%
\usepackage{amsthm}%
\usepackage{mathrsfs}%
\usepackage{xcolor}%
\usepackage{textcomp}%
\usepackage{booktabs}%
\usepackage{siunitx}
\usepackage{hyperref}
\begin{document}

\title{Crosstalk-mitigated microelectronic control for optically-active spins}

\author{Hao-Cheng Weng}
\email{haocheng.weng@bristol.ac.uk}
\affiliation{Quantum Engineering Technology Labs, H. H. Wills Physics Laboratory and Department of Electrical and Electronic Engineering, University of Bristol, Bristol BS8 1TL, United Kingdom}

\author{John G. Rarity}
\affiliation{Quantum Engineering Technology Labs, H. H. Wills Physics Laboratory and Department of Electrical and Electronic Engineering, University of Bristol, Bristol BS8 1TL, United Kingdom}

\author{Krishna C. Balram}
\affiliation{Quantum Engineering Technology Labs, H. H. Wills Physics Laboratory and Department of Electrical and Electronic Engineering, University of Bristol, Bristol BS8 1TL, United Kingdom}

\author{Joe A. Smith}
\email{joe.a.smith@sheffield.ac.uk}
\affiliation{School of Electrical and Electronic Engineering, University of Sheffield, Sheffield S1 3JD, United Kingdom}
\affiliation{Quantum Engineering Technology Labs, H. H. Wills Physics Laboratory and Department of Electrical and Electronic Engineering, University of Bristol, Bristol BS8 1TL, United Kingdom}

\date{\today}

\begin{abstract}
To exploit the sub-nanometre dimensions of qubits for large-scale quantum information processing, corresponding control architectures require both energy and space efficiency, with the on-chip footprint of unit-cell electronics ideally micron-scale. However, the spin coherence of qubits in close packing is severely deteriorated by microwave crosstalk from neighbouring control sites. Here, we present a crosstalk-mitigation scheme using foundry microelectronics, to address solid-state spins at sub-100 \textmu m spacing without the need for qubit-detuning. Using nitrogen-vacancy centres in nanodiamonds as qubit prototypes, we first demonstrate 10 MHz Rabi oscillation at milliwatts of microwave power. Implementing the active cancellation, we then prove that the crosstalk field from neighbouring lattice sites can be reduced to undetectable levels. We finally extend the scheme to show increased qubit control, or effectively, the spin coherence under crosstalk mitigation. Compatible with integrated optics, our results present a step towards scalable control across quantum platforms using silicon microelectronics.
\end{abstract}


\maketitle

\section{Introduction}

The field of quantum computing, underscored by demonstrations of computational quantum advantages \cite{arute2019quantum,zhong2020quantum}, has shifted towards engineering scalable systems, both in terms of qubit quality  (coherent gate operations \cite{evered2023high}) and quantity ($>$ 1 million qubits) \cite{kjaergaard2020superconducting, fowler2012surface}. 

To scale up, the key advantage of solid-state systems is the prospect of achieving close packing of qubits considering energy efficiency and space constraints in cryogenic operations. However, for large-scale systems, performance becomes limited by classical control and wiring \cite{gambetta2020ibm, bravyi2022future}, especially when close packing brings with it the problem of qubit crosstalk between neighbouring controls. Without microwave crosstalk mitigation, two identical qubits have to be separated by millimetres, which prohibits placing more than a few qubits on the same chip (5 mm by 5mm) and scalability in any sense. 

A variety of control signals for quantum systems fall in the megahertz to gigahertz-frequency domain \cite{kurizki2015quantum, riste2020microwave}. These systems range from superconducting transmon qubits to spin qubits in solid-state systems and trapped ions \cite{bardin2021microwaves,brecht2016multilayer, lekitsch2017blueprint} where two-level systems are addressed by microwave pulses. Progress on crosstalk mitigation has been made particularly in superconducting qubits \cite{spring2022high}, however these designs are not directly adaptable to qubits with optical access such as colour centres in diamond, 2D materials, and trapped ions. Close proximity of metallic structures to optics enhances scattering and loss of photons, posing a significant challenge. 

Previous work in trapped ions \cite{piltz2014trapped} employed gradient fields to give unique frequency addressability to each qubit, preventing crosstalk. For solid-state defects, recent work proposed alternative spin driving schemes with strain-tunable electrodes \cite{wang2023field} and electromechanical control \cite{clark2024nanoelectromechanical} to achieve localised control fields. Nevertheless, requiring distinct microwave frequencies per qubit is experimentally costly and mechanical control is challenging to fabricate.

In this work, we present a new design for site-defined control with silicon microelectronics, prototyping with foundry Application-Specific Integrated Circuits (ASIC) and nitrogen-vacancy centres in diamonds (NVs). 
NVs are well-characterised spin qubits with optical interfaces \cite{pompili2021realization, abobeih2022fault, bian2021nanoscale}. Moreover, the room-temperature operation of NV spin systems and point-like localisation makes them ideal prototypes for testing large-scale control architectures in the gigahertz regime. 

The complexity of microelectronics allows fabricable quantum control systems at sub-mm scale \cite{kim2019cmos}, as compared to macroscale metallisation typically realised with printec circuit boards \cite{yang2019microstrip}. We introduce a dual-loop transducer ASIC design which effectively isolates single control sites at sub-$\SI{100}{\micro\meter}$ qubit packing densities. Our result illustrates the possibility of scalable control for spins individually and simultaneously in a compact device. The scheme is scalable and paves the way for further combination with active microelectronics and integrated optics.

\begin{figure*}[t]
\centering
\includegraphics[width=0.8\textwidth]{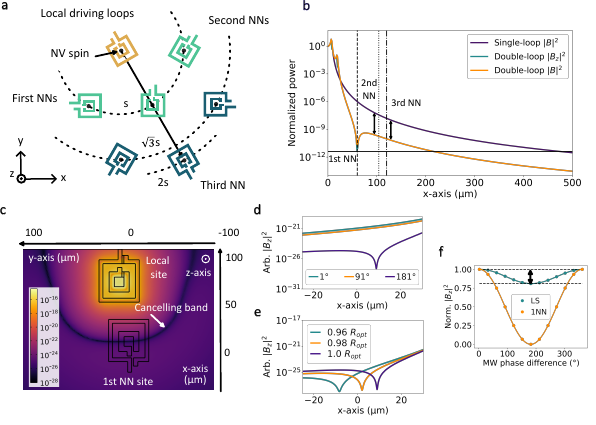}
\caption{\textbf{Active crosstalk cancellation scheme}. \textbf{a} Hexagonal packing of driving sites: to the upper-left (golden) site, the sites in light green are first Nearest Neighbour (NN) sites at a distance of $s$. The second (in dark green) and third NN sites (in purple) are $\sqrt{3}s$ and $2s$ away, respectively. Each site is presented by two concentric loops hosting a spin qubit (black dot in the centre) set in the x-y plane, where the outer loop produces a cancelling signal out-of-phase with the inner loop signal. 
\textbf{b} Comparison of the magnetic field power at different distance from the local driving site, showing field cancellation. See analytical simulation method in Appendix A. The power attenuation at second and third NN sites is labelled. \textbf{c} Finite-element simulation of this structure realised following foundry microelectronics design rules. The outer loops are $\SI{38}{\micro\meter}$ in diameter and the inner loops $\SI{15.5}{\micro\meter}$. The power of the out-of-plane magnetic field component $|B_z|^2$ is plotted.  The magnetic field power plot has units in Tesla squared, when driven at 1 volt by a 50 ohm coaxial source. For \textbf{d-e}, the same unit applies. \textbf{d} The field cancellation shown precisely when the phase difference (lines in graph) is at 181$^\circ$ in the simulation. \textbf{e} The point of maximal cancellation spatially tuned by changing the driving ratio $R_{opt}$ between the two loops. \textbf{f} Comparison of the driving power $|B_z|^2$ at the local site (LS) and the first NN site (1NN), in a sweep of set microwave phase difference between the inner and outer loop.}\label{fig1}
\end{figure*}

\section{Active crosstalk cancellation}\label{sec_idea}

Spin-based quantum systems are commonly electrically driven by a close metallic structure to ensure efficient and localised manipulation. However, for optically-active spins, this structure should be located at least several optical wavelengths away from the spin, to prevent optical scattering and loss in the photonic channel. Therefore, there is a trade-off between the deliverable localised field at the spin and the need to reduce optical loss by keeping the metallic structure at some distance from the dipole-like emission. To account for this, we introduce a square loop geometry (in Fig. \ref{fig1}a) for dedicated spin control of each site. This maximises the area into which photonic structures can be introduced, whilst also maximising the contact area of the driving transducer for a concentrated and directed field. 

On the other hand, as microwave controls for spin systems have a near infinite wavelength $\lambda > \SI{1}{\centi\meter}$, compared to micrometre features, their spatial extent can be significant, deleteriously affecting secondary systems located nearby. For a single loop delivering 0 dB power to the control site at the origin, a significant power (-60dB or 0.5\% reduction of the spin coherence time, Appendix A) remains at $\SI{60}{\micro\meter}$ from the centre (purple line in Fig. \ref{fig1}b). Common methods to remove the residual field, such as enclosing Faraday cages or lead shielding, are incompatible with planar fabrication processes.

However, by adopting multi-layer metallisation from silicon microelectronics, we have the flexibility to route signals out of the plane, enabling complex geometries. In this work, we exploit this by considering two lithographically-defined concentric loops, as shown in Fig. \ref{fig1}a. By driving the two loops 180$^\circ$ out-of-phase, the two counter-propagating fields cancel out, away from the local control site. With finite-width current carrying wires and a rectangular geometry, we examine the cancellation in a fabricable device through a gigahertz finite-element simulation (Appendix B). Note that the square shape of the loops is constrained by foundry design rules for the metal layer and works similarly to circular loops (Appendix A). In Fig. \ref{fig1}c, we show the magnetic field power ($|B_z|^2$) on this device containing a local driving site and a first NN site. By setting the driving current ratio and phase difference, the maximally cancelled region forms a band. In Fig. \ref{fig1}d, by tuning the relative phase between the feeds driving each loop, complete cancellation happens at 181$^\circ$, producing -50 dB attenuation at the first NN (x=\SI{10}{\micro\meter} following the coordinates in Fig. \ref{fig1}c). The cancelling point can be fine tuned spatially by changing the driving ratio between the two loops. In Fig. \ref{fig1}e, by sweeping the driving voltage ratio between the inner and outer loops at the local site $R$ around $R_{opt}$, the optimised ratio such that cancelling happens at $\textrm{x}=\SI{10}{\micro\meter}$, the dip can be tuned between $\textrm{x}=0$ to $\SI{20}{\micro\meter}$. In Fig. \ref{fig1}f, we compare the field at the local site centre (top of Fig. \ref{fig1}c) and the remote site (bottom). Along the phase sweep, only a slight reduction in power ($<20\%$) is observed at the local site. This allows the cancelling scheme to operate without significantly sacrificing the local driving field. In practice, once the cancellation condition is found, the microwave source amplitude can be scaled to achieve a desired driving power.

This design allows (i) a much more localised field and (ii) precisely cancelling of the field at a distance corresponding to the first Nearest Neighbour (NN) sites. To see how the the two-loop scheme works, we compare the total magnetic power $|B|^2$ (normalised to the control site at $\textrm{x}=0$) at NN sites spaced with hexagonal packing (Fig. \ref{fig1}b orange line). The field cancels perfectly at the first NN site ($\textrm{x}=\SI{60}{\micro\meter}$), showing high power extinction (-120 dB, when 0 dB is delivered at $\textrm{x}=0$), or equivalently an additional -50 dB attenuation compared to the single-loop case. As the dominant field component is in the z-direction, the two-loop $|B|^2$ extinction ratio is similar to that of $|B_z|^2$. At the second and third NN sites ($\textrm{x}=\SI{104}{\micro\meter}$ and $\textrm{x}=\SI{120}{\micro\meter}$), as the field decays faster in the two-loop scheme (following a $1/x^{15}$ power law decay as compared to the $1/x^{6}$ for the single-loop scheme when fitted), an additional -20 dB power attenuation is applied. The scheme scales for all NN sites at the same distance from the local site.

\section{Hybrid device realisation}\label{sec_fab}

\begin{figure}[t]
\centering
\includegraphics[width=\columnwidth]{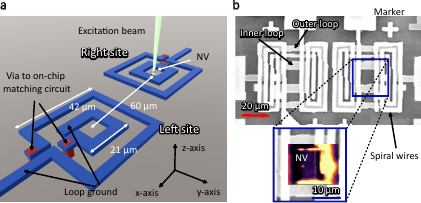}
\caption{\textbf{Hybrid microelectronic device.} \textbf{a} Schematic of the hybrid device showing the dual-loop structures realised on an ASIC, combined with an Nitrogen-Vacancy centre (NV) in a nanodiamond. The dimensions of inner/outer loops follow that simulated in Fig. \ref{fig1}c. The right site shows the NV in the positioned nanodiamond addressed by the excitation laser. See details in Appendix C. For each loop, one end is connected to ground and the other end connected to the on-chip matching circuit through the red vias (see Appendix C). \textbf{b} An exemplar scanning electron microscope image of the device. Markers are used during the positioning process for alignment. The loop structures are underneath spiral wires not used in the current experiment. The inset shows a confocal scan of the optically-active NV (shown by the central Gaussian spot) in the centre of the structure.}\label{fig2}
\end{figure}

To demonstrate the active crosstalk cancellation in a scalable platform, we leverage a commercial $\SI{0.13}{\micro\meter}$ back-end-of-line silicon foundry process with aluminium layers, isolated with silicon dioxide, and tungsten vias (see Appendix D). Unlike PCBs (Printed Circuit Boards), a millimetre scale technology constrained by the use of plated holes to form through plane vias, a silicon foundry process allows us to achieve micron scale features. In Fig. \ref{fig2}a, the structure is realised in two rectangular loops of the same dimensions, following the gigahertz finite-element simulation result in Fig. \ref{fig1}c-f. As a demonstration, nanodiamonds containing single optically-accessible NVs are lithographically positioned in a post-processing step (see Appendix C). The structure imaged using scanning electron microscopy is shown in Fig. \ref{fig2}b. The hybrid device is studied with confocal microscopy under 515 nm laser excitation to identify an isolated NV at the control site (Fig. \ref{fig2}b inset) through photoluminescence (PL).

\begin{figure*}[t]
\centering
\includegraphics[width=0.9\textwidth]{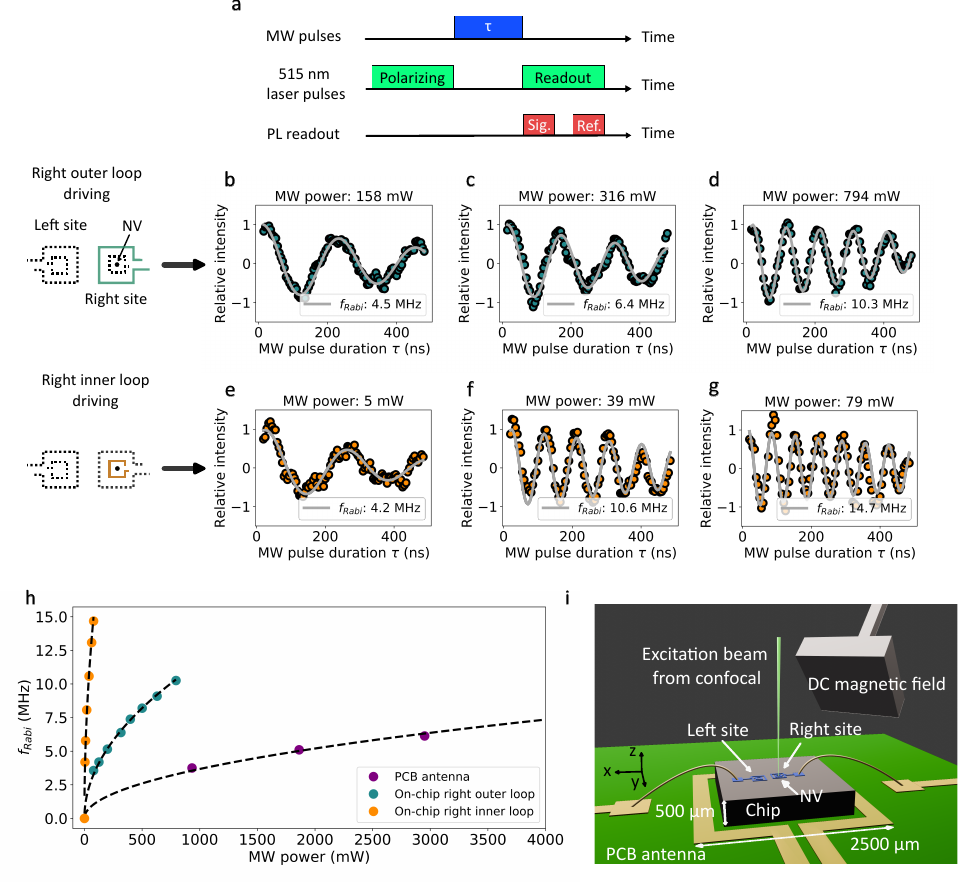}
\caption{\textbf{On-chip electron spin driving.} \textbf{a} Pulse sequences for Rabi oscillation experiments. The first part of the photoluminescence (PL) readout is the signal (Sig.) that contributes to the state-dependent intensity contrast and the second part is a reference (Ref.) for normalisation of counts. The metallic loops are studied by performing Rabi oscillations on the same NV electron spin. The spin associated with the isolated NV is driven by the on-chip right outer loop (\textbf{b-d} in green) and the on-chip right inner loop (\textbf{e-g} in orange) showing the Rabi oscillation at different driving microwave (MW) power. Note that where the geometry is shown by dashed line means that the structure is not driven. The applied microwave power and Rabi frequency are labelled. Here, the microwave power is normalised to the loop $S_{11}$ efficiency, i.e., the microwave reflection coefficient (see Appendix D), representing the power delivered by the loop structures. \textbf{h} Comparing the on-chip loop structures to the global PCB antenna demonstrates an observed two-hundred fold improvement in driving efficiency above 10 MHz. \textbf{i} A schematic diagram showing the spatial relations between the on-chip loops and the PCB antenna. Dimensions are not to scale.}\label{fig3}
\end{figure*}

The on-chip loops (with on-chip matching circuitry, see Appendix C) are then connected to external electronics and microwave sources (see Appendix D). By controlling the NV in Fig. \ref{fig2}b, we can evaluate the microelectronics performance for manipulating a quantum system. We perform Rabi oscillations on the NV electron spin (see Appendix D) by applying microwave and laser pulses presented in Fig. \ref{fig3}a. As the microwave pulse width $\tau$ is swept, the oscillation of PL intensity reveals the spin state transition at a frequency $f_{Rabi}$ proportional to the square root of microwave power applied. We compare the Rabi oscillation of the same enclosed NV when driven with the on-chip right outer loops (Fig. \ref{fig3}b-d) and right inner loops (Fig. \ref{fig3}e-g), following the labels in Fig. \ref{fig2}a.

In Fig. \ref{fig3}h, plotting the power dependence of the observed Rabi frequency indicates that on-chip driving (compared to driving with global PCB antenna, see Appendix D) is significantly more efficient, as expected from the micron proximity to the radiation source (following the $1/r^3$ power law decay where $r$ is distance from the source), as opposed to approximately \SI{500}{\micro\meter} for the external antenna (Fig. \ref{fig3}i). The inner loop also drives more efficiently than the outer loop. Specifically, our result shows that the on-chip metallic loop interface can be two hundred times more efficient, achieving a Rabi frequency $f_{Rabi}\simeq$10 MHz with $<$40 mW (power at the inner loop). This is fast enough for coherent control of the NV electron spin considering $1/f_{Rabi} < T_2/10$ where the typical $T_2$ coherence time for NV electron spin in nanodiamond is around \SI{1}{\micro\second} \cite{knowles2014observing}. It is worth noting that more efficient driving has been achieved by identifying NVs close to patterned metallisation on diamond \cite{mariani2020system} or nearby wires \cite{wang2015high}. However, our demonstration presents a scalable solution with foundry microelectronics, providing both the uniformity of field in the loop to favour NV positioning and compatibility with integrated photonics.

\section{Crosstalk-mitigated spin control}\label{sec_result}

With the ability to efficiently manipulate the spin, we progress to demonstrate the crosstalk mitigation scheme. By operating both the inner and outer loops (left site in Fig. \ref{fig2}a) at the site adjacent to the isolated spin site, our aim is to observe the cancellation of the driving field from this neighbouring control site, at the position of the isolated spin. 

The cancellation condition can be found experimentally by first balancing the effect of the driving fields from the inner and outer loops on the NV spin. The driving field strength can be inferred from continuous-wave Optically Detected Magnetic Resonance (ODMR) measurements (see Appendix D). When the microwave source frequency is swept, a dip with a certain depth, the ODMR contrast, is observed at the resonance of the spin state transition. In the low pump regime (small microwave power) \cite{dreau2011avoiding}, this contrast is proportional to the microwave power. This linearity demonstrates that the ODMR contrast can be directly used to quantify and balance the microwave power measured at the NV site.

\begin{figure*}[hp!]
\centering
\includegraphics[width=0.8\textwidth]{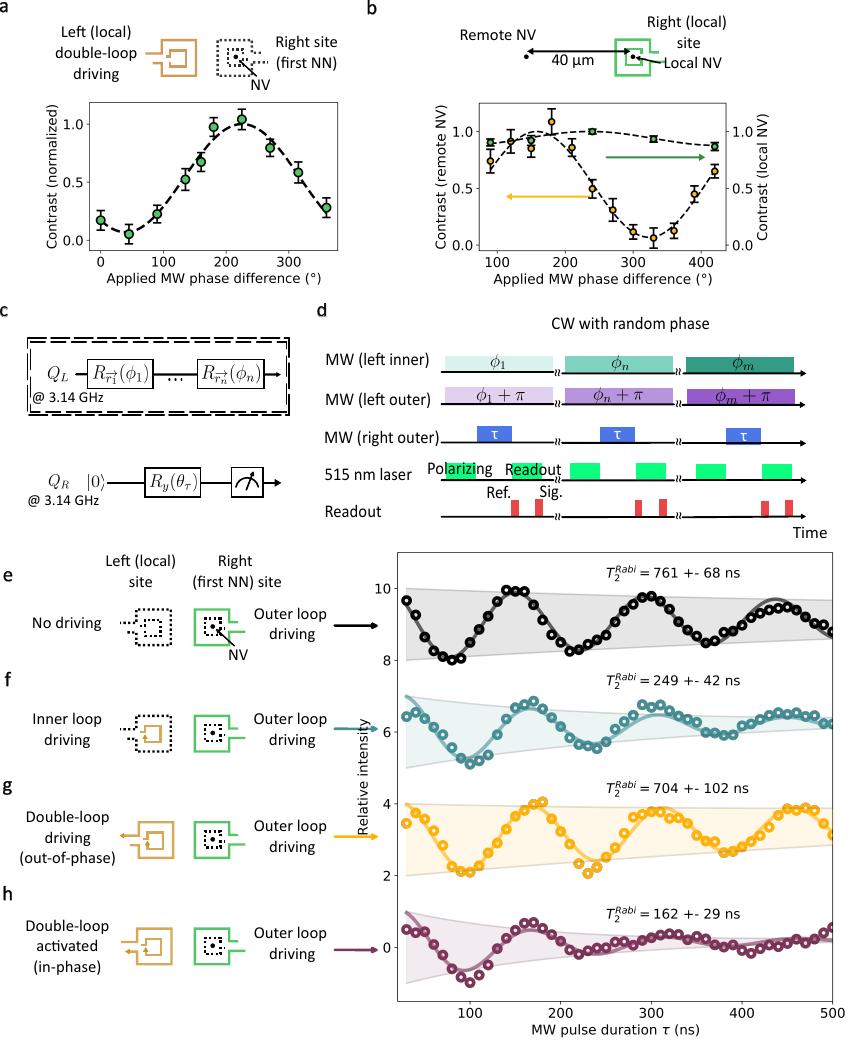}
\caption{\textbf{Active crosstalk cancelling}. \textbf{a} Under the two-loop driving, a sinusoidal relation between the ODMR contrast and the applied phase difference between the two loop feeds is observed. \textbf{b} Local NV control is minimally affected under the field cancellation protocol. Orange data points shows results for a remote NV and green data points for the local NV in the right site. The remote NV 40 \textmu m away is chosen for experimental accessibility. Both the remote and local NV contrast are fit to a sinusoidal function and their relative phase shift is due to difference of NV orientation. The set phase is different from \textbf{a} due to the difference in experimental conditions. The electron spin resonance is not shifted here to make sure the the spins are driven under the same condition. \textbf{c} Gate sequences for the crosstalk-mitigated spin manipulation, considering two qubits $Q_L$ and $Q_R$. Random qubit rotations $R_{\vec{r_n}}(\phi_n)$ are performed on $Q_L$ and $R_y(\theta_\tau)$ (rotation around the y-axis on the qubit Bloch sphere) performed on $Q_R$ before state readout. There is no detuning between $Q_R$ and $Q_L$- both qubits operate at 3.14 GHz. \textbf{d} Corresponding pulse sequences for demonstrating the crosstalk-mitigated spin manipulation. Microwave pulses in different color shade (top two rows) corresponds to different phases. Here, the random phase change happens every one second (while the inner and outer loop is always out of phase), over the $>$ 3 hour accumulation time for each experiment. The Rabi sequence follows that performed in Fig. \ref{fig3}a. The microwave power applied to the left loops are larger than that to the right loop, allowing a clear observation of spin decoherence under noise. \textbf{e} Rabi oscillations driving only the right outer loop. The data is presented with a sinusoidal decay fitting and corresponding $T_2^{Rabi}$. \textbf{f} Rabi oscillations with the left inner loop also driving, showing reduced control of spin qubit. \textbf{g} Rabi oscillations protected by the two-loop active cancelling scheme (driving out of phase), showing regained control. \textbf{h} Rabi oscillations when the two left loops are in phase, showing increased crosstalk. Note that the Rabi oscillation for \textbf{e-h} has the same frequency around 7 MHz.} \label{fig4}
\end{figure*}

Once the driving fields are balanced, with both loops activated, one needs to identify the set phase between the microwaves applied to the loops. In Fig. \ref{fig4}a, we observe the sinusoidal relation between the applied phase difference and the ODMR contrast, as expected in Fig. \ref{fig1}f. The cancelling phase condition is then the destructive interference point ($45^\circ$) where the contrast approaches its minimum. From the fitting in Fig. \ref{fig4}a, the minimum point shows a normalised power of $0.06\pm0.06$, meaning that the microwave power is undetectable within error bars. From our simulation (Fig. \ref{fig1}f), the field cancellation for a remote NV only minimally affect the driving of a local NV. This is demonstrated experimentally (Fig. \ref{fig4}b) when the driving amplitude is set for a remote NV positioned 40 \textmu m away. Minimal change ($<15\%$) is observed on the local NV, compared to the almost unity modulation of the remote NV over the set phase sweep. In addition, we observe minimal change in the Rabi driving efficiency of the local NV under the active cancellation operation.

Under this crosstalk-mitigation condition, we can now demonstrate coherent spin control. Our scheme considers two adjacent control sites with associated qubits, $Q_L$ and $Q_R$, shown in Fig. \ref{fig4}c. Microwave control for $Q_L$ is allowed to perform arbitrary single qubit rotations on the spin two-level system, whilst independently, $Q_R$ is coherently controlled and evaluated. The crosstalk-mitigated control of $Q_R$ (by cancelling the crosstalk field from the other site) means that the control of $Q_R$ will remain coherent, independent of the control signal for $Q_L$. Otherwise, the microwave crosstalk would result in random rotations on $Q_L$, diminishing the ability to perform coherent qubit control. 

For $Q_L$, the arbitrary rotations $R_{\vec{r_n}}(\phi_n)$ corresponds to resonant microwave pulses of different phases, starting time, and end time. Randomly chosen rotations can thus be simulated by a resonant microwave signal (at 3.14 GHz) in continuous wave (considering that random pulsing would average out in time) with randomly changing phases. We deliver the same signal to the left inner and outer loop (labelled in Fig. \ref{fig2}a) with the $\pi$ phase shift for the cancelling condition to hold. Pulse sequences are shown in Fig. \ref{fig4}d. Rabi sequences (microwave pulses of variable length $\tau$) are then delivered to the right outer loop (similar to Fig. \ref{fig3}a).

With the microwave signal in Fig. \ref{fig4}d applied, we compare the quantum control of $Q_R$ under different conditions. Fig. \ref{fig4}e shows the Rabi oscillation performed by the right outer loop alone, serving as a reference. A Rabi frequency of 7 MHz is observed with a fitted decay time $T_2^{Rabi}=761\pm 69$ ns, within which the spin can be coherently driven. In this case, $T_2^{Rabi}$ is an effective timescale set by the Rabi pulses applied, intrinsic spin dephasing, and spin energy relaxation in the host environment, characterising the controllability of the spin \cite{hanson2006room}. In Fig. \ref{fig4}f, when the left inner loop is turned on, crosstalk noise is introduced to the Rabi oscillation. The random noises, averaged over the experimental accumulation time, effectively leads to decreased contrast in the Rabi oscillation and thus a decreased $T_2^{Rabi}$ to $249\pm 42$ ns. The large decrease in $T_2^{Rabi}$ is expected under strong random driving noise and matches simulations. 

However, we observe in Fig. \ref{fig4}g that when both the left inner and outer loop are on and operated out-of-phase, the Rabi oscillation is protected, with $T_2^{Rabi}$ tripled in duration to $704\pm 103$ ns, meaning that the coherence of the qubit control returns to the case without the neighbouring drive field (within the range of error bars). From the recovery of the $T_2^{Rabi}$ value, one can estimate the extinction ratio of the crosstalk microwave power to be 0.03 (-15 dB). 

In comparison, in Fig. \ref{fig4}h, when the signals in the left inner and outer loops are in phase, the noise is considerably increased, with $T_2^{Rabi} = 162\pm 30$ ns. Note that $T_2^{Rabi}$ here only suggests a general idea of spin controllability. The spin coherence, in terms of $T_2^{*}$, $T_2$, and $T_1$, is also protected to different level under our crosstalk mitigation scheme. 

It is worth noting that we look at the coherence of the same spin with its associated spin environment in the presence of a neighbouring qubit and show that it suffers a decrease in coherence when the neighbouring qubit is driven, unless using the cross-talk mitigation scheme. Our results show that the crosstalk mitigation effectively removes the noise generated by the neighbouring control line. The same level of noise attenuation in our case would require a qubit detuning of 40 MHz where a finite-size frequency band for each qubit imposes architectural complexity, requiring generation of large gradient field at scale \cite{jakobi2017measuring}.

\begin{figure*}[t]
\centering
\includegraphics[width=\textwidth]{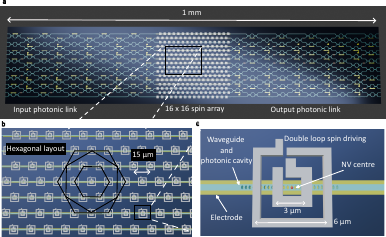}
\caption{\textbf{Schematic device for large-scale quantum information processing: spin qubit integrated with crosstalk-mitigated microelectronics and integrated photonics.} \textbf{a} The schematic device. The 16 by 16 spin qubit array are interconnected by photonic links on both side. The photonics can be used for routing of photons and photonic gate that entangles the optically-active spins. \textbf{b} A zoomed in image of the spin qubit arrays. The qubit are layout following the proposed hexagonal architecture in Fig. \ref{fig1}a with inter qubit spacing of 15 \textmu m. \textbf{c} A zoomed in image of a qubit site. The optically-active spin qubit (taking NV centre for example) is coupled to the photonic cavity on a strip waveguide. NV photonic resonances can be electrically tuned with the yellow electrode under the waveguide. The spin is locally manipulated by the double loop driving. Here the inner loop and outer loop are 3 \textmu m and 6 \textmu m in diameter respectively.} \label{fig5}
\end{figure*}

\section{Conclusions}\label{sec12}

We have demonstrated a site-defined control architecture using foundry microelectronics, for the scalable manipulation of spin qubits. With NVs lithographically positioned at predefined sites, we showed that their associated electron spin can be driven efficiently, important for quantum computing, which requires liquid helium operation to extract coherent photons from NVs, and hence has limited thermal budget. Most importantly, our results demonstrate that crosstalk in driving adjacent spins can be strongly mitigated through an active cancellation scheme, without significantly sacrificing this driving efficiency. The scheme works without frequency detuning spin qubit transitions \cite{neumann2010quantum, jakobi2017measuring, sekiguchi2022optically}, allowing distinct control of sets of identical qubits. Note that our crosstalk mitigation method through hardware design prevents the time-consuming pair-wise calibration and sophisticated signal compensation typically used in addressing qubit systems \cite{acharya2024quantum}. The extinction ratio of the field cancellation can also be improved by more precisely balancing the driving of the inner and outer loops (see Appendix D). In this work, nanodiamonds are primarily used for ease of prototyping. For large scale quantum information applications, use of implanted NVs in strain-minimised diamond membrane structure \cite{guo2024direct} would minimise difference in local NV properties which could also be locally tuned \cite{dolde2011electric}.

Fig. \ref{fig5} showcases how the double-loop crosstalk-mitigated microelectronics would fit into existing large-scale quantum information processing hardware for optically active spins \cite{simmons2024scalable,li2024heterogeneous, wan2020large}. The micron-scale separation of spin defects in the architecture highlights the need of crosstalk-mitigation microelectronics. Fig. \ref{fig5}a shows the integration of a 16 by 16 spin array with input and output photonic links for selective addressing and entangling photonic gates. Notably the photonic resonant transition of NVs can be tuned electrically \cite{li2024heterogeneous}, which allows single spins along the waveguide to be selected and addressed separately \cite{robledo2011high}. The tunability also enables multi-qubit gate or entanglement of spins across waveguides through interference of emitted photons \cite{beukers2024remote} of the same wavelength. The scheme could be readily adopted to previous work \cite{li2015coherent, wan2020large,li2024heterogeneous}. Technological challenges remain in realising a platform that combines a visible wavelength photonics with multilayer microelectronics, although this has seen dramatic progress in recent years with silicon nitride integration \cite{sacher2018monolithically, alexander2024manufacturable}, and the subsequent integration of the thin film diamond containing the qubits \cite{wan2020large}. With qubit spacing of 15 \textmu m and the inner(outer) loop dimension of 3(6) \textmu m, the whole hundred-physical-qubit device has a dimension $<$ 1mm$^2$, demonstrating high level of integration. It is also possible to pursue higher densities at smaller process nodes by exploiting the presented crosstalk cancellation scheme and significant headroom in the current capacity of the existing circuit. Considering NV positioning accuracy of 50 nm and a sub-micron optical modes through confocal microscopy or waveguide mode excitation \cite{bhaskar2017quantum}, sub-micron separation should be achievable. 

Foundry microelectronics also gives access to mature silicon device libraries to realise static field tuning \cite{arai2015fourier}, electrical spin readout \cite{bourgeois2015photoelectric, gulka2021room}, photon detection, and logical electronics \cite{kim2019cmos} in the same platform. In combination with photonic integrated circuits, a large-scale spin-photon interfaced quantum information processor could be realised, overcoming space and energy constraints. The scheme presented here is also translatable to alternative quantum platforms that use microwave driving, including other colour centres in diamond \cite{castelletto2020silicon}, single spin molecules \cite{gaita2019molecular}, quantum dots \cite{lawrie2023simultaneous} or superconducting circuits \cite{mitchell2021hardware}.

%\section{Data availability}
%The data that support the plots within this paper and other findings of this study are available from the corresponding author upon reasonable request.

%\section{Code availability}
%The code used to generate the plots within this paper is available from the corresponding author upon reasonable request.

\begin{acknowledgments}
The authors thank J. Monroy-Ruz for building the confocal microscope used in this experiment and S. Komrakova, J. Frazer, and F. Malik for fabrication support. The authors thank J. Monroy-Ruz and J. Matthews for useful discussions. The authors acknowledge funding support from the Quantum Computing and Simulation Hub Partnership Resource Project RFSQ and the Engineering and Physical Sciences Research Council (EPSRC) grant QC:SCALE EP/W006685/1.
\end{acknowledgments}

%Appendices for detailed supplementary information
\appendix

\section{Analytical model expression}

The analytical model presented (Fig. \ref{fig1} in the main text) uses the static Biot-Savart approximation. We consider the off-axis field of a current loop centred at the origin in the x-y plane. For any point in the space, the magnetic field is defined in the cylindrical coordinate system with the component in axial direction
\begin{equation}
    B_z(r,z)=\gamma\left(E(k)\frac{1-\alpha^2-\beta^2}{(1+\alpha)^2+\beta^2-4\alpha}+K(k)\right)
\end{equation}
and in the radial direction
\begin{equation}
    B_r(r,z)=\gamma\left(E(k)\frac{1+\alpha^2+\beta^2}{(1+\alpha)^2+\beta^2-4\alpha}-K(k)\right),
\end{equation}
where $r=\sqrt{x^2+y^2}$ and $\gamma = \frac{i\mu_0}{2a\pi\sqrt{(1+\alpha)^2+\beta^2}} $. Here, $\mu_0$ is the permeability constant, $a$ is the radius of loop, $i$ is the current flow in the loop wire, $K(k)$ is the elliptic integral function of first kind, $E(k)$ is the elliptic integral function of second kind, and constants defined as $\alpha=r/a$, $\beta=z/a$, $\gamma=z/r$, and $k=\sqrt{\frac{4\alpha}{(1+\alpha)^2+\beta^2}}$.

\section{Finite-element simulation details}

To verify the device design, a finite-element model of the structure is carried out in COMSOL Multiphysics following the analytical model. The double-loop structure is constructed with aluminium metallic loops surrounded by silicon oxide with relative permittivity $\epsilon_r=4.1$. Lumped ports are defined between the input and output metallic wire of the loops and excited by a voltage source with the port phase set in simulation. The Electromagnetic wave, Frequency domain analysis COMSOL package is used to study the device at 2.87 GHz. The minor difference between analytical and simulation results is due to the physical width of the wires of loop structure that is not considered in the analytical model.

\section{Device fabrication details}

We employ the $\SI{0.13}{\micro\meter}$ IHP-SG13G2 BiCMOS technology with two thick metal layers for current-carrying applications and five thin metal layers for interconnections. The track width of the inner (outer) loops are $\SI{5.5}{\micro\meter}$ ($\SI{4}{\micro\meter}$), both on the same $\SI{2}{\micro\meter}$ thick aluminium layer. 

For nanodiamond positioning, nanodiamonds are sized around $\SI{50}{\nano\meter}$ diameter, milled from high pressure high temperature diamond (Nabond). First, a post-process inductively coupled plasma etch is used to release the inner region of the device, creating a $\SI{5}{\micro\meter}$-deep trench. This allows nanodiamonds to be positioned at the centre of the loops. A layer of polymethyl methacrylate (PMMA) is then coated and patterned by electron beam lithography (Raith Voyager). This creates sites on the PMMA mask on which the nanodiamond (suspended in methanol) is then dropcast. Removing the PMMA then results in nanodiamonds positioned precisely on the predefined sites.

\section{Experimental setup}

A scanning confocal microscope is used to optically address the NV in this work. With a 0.9 numerical aperture microscope objective (Olympus M Plan N 100x/0.90NA), the excitation beam (515 nm, Cobolt 06-MLD) is focused on the sample at a nearly diffraction-limited spot ($<$ \SI{1}{\micro\meter} diameter). The NV photoluminescence is collected through the same lens and separated from the excitation path using a dichroic mirror. By using a high NA objective, the NV PL is extracted from the background fluorescence such as scattering from metal layers. Electronically, for the active cancelling demonstration, the microwave source (Rohde \& Schwarz SMB100AP20) is split by a power splitter (Mini-Circuits ZN2PD2-63-S+) for driving the inner and outer right metallic loop interface. On one path, an additional tunable power attenuator (Mini-Circuits RCDAT-6G-120H) and phase shifter (Vaunix LPS-802) are applied. Both paths are amplified (Microwave Amplifiers Ltd AM4-2-6-43-43R) and fed to an isolator (DiTom Microwave D3C2040) before the device-under-test. Timetagging and pulsing are recorded by UQDevices Logic-16 and Swabian Instruments Pulse Streamer 8/2. The electronic setup for crosstalk-mitigated spin control only differs with another random phase shifter applied before power splitting. Pulsing is achieved using a microwave switch (Mini-Circuits, ZASW-2-50DR+, DC-5 GHz) in Rabi oscillation measurements.

We perform electron spin Rabi oscillations by driving the NV ground state triplets ($|0\rangle_{e}$ and $|\pm1\rangle_{e}$). Fig. \ref{fig3}a shows the corresponding pulse sequences. The electron spin is first polarised into the $|0\rangle_{e}$ state by a $\SI{3}{\micro\second}$ long 515 nm laser pulse. A microwave pulse of length $\tau$ is then applied to resonantly drive the transition between $|0\rangle_{e} \longleftrightarrow |+1\rangle_{e}$ (with the ground state zero-field splitting around 2.87 GHz). In our measurement, an external static magnetic field is applied using a strong permanent magnet in order to lift the degeneracy of the two transitions ($|0\rangle_{e} \longleftrightarrow |+1\rangle_{e}$ and $|0\rangle_{e} \longleftrightarrow |-1\rangle_{e}$) and to shift the resonances such that it is best matched to the on-chip loop resonance. Another laser pulse of $\SI{3}{\micro\second}$ is used to readout the spin state by the PL intensity (stronger in the $|0\rangle_{e}$ state and weaker in the $|\pm 1\rangle_{e}$ state due to state-dependent intersystem crossing). The PL intensity is extracted by summing the photon count in the first $\SI{0.5}{\micro\second}$ interval and normalising to the count summed over the last $\SI{1.5}{\micro\second}$. We use the same laser pulse to polarise and readout the electron state. In Fig. \ref{fig3}b-g, the Rabi measurements are presented (in the y-axis) with the relative intensity normalised between $\pm1$ for comparison. For the Rabi oscillation results presented in Fig. \ref{fig4}e-h, the error in decay time is deduced from the variance of the parameter estimation during fitting, presenting one standard deviation error. The slight drift in oscillation is an artifact due to device heating under microwave power.

In the continuous-wave ODMR measurement, the 515 nm laser is constantly on while the frequency of the microwave source is swept around the resonance of electron spin transitions. Again, due to the difference of PL intensity for the electron states, a dip is observed at resonances with a certain depth (the ODMR contrast). The contrast is extracted by fitting the measured PL intensity across frequency sweeps to a Gaussian dip. Error bars, showing one standard deviation in photon counts, arise from the photon shot noise and low drift of the experimental setup.

The field cancellation extinction ratio (directly proportional to the residual crosstalk field) is undetectable (within experimental error bar) from the CW-ODMR measurements and Rabi oscillation lifetimes (Fig. \ref{fig4}e-h) directly. Fitting the results in Fig. \ref{fig4}a, a -15 dB extinction ratio of crosstalk field power is inferred. The extinction ratio can be further improved by precisely balancing the driving of the inner-outer loops to the optimal ratio. For our case demonstrated in Fig. \ref{fig4}(e-h), a -20 dB imbalance of microwave power (from the left loops) at the NV would diminish $T_2^{Rabi}$ only by 2\% under the active cancelling protection and a -30 dB imbalance for 0.2\%.

\bibliography{bibliography}

\end{document}